\documentclass[supplement]{ptptex}

\usepackage{graphicx}

\title{
Nonextensive/dissipative correspondence in relativistic hydrodynamics%
}

\author{
Takeshi \textsc{Osada}$^{1,}$\footnote{E-mail: osada@ph.ns.musashi-tech.ac.jp}
and Grzegorz \textsc{Wilk}$^{2,}$\footnote{E-mail: Grzegorz.Wilk@fuw.edu.pl}%
}

\inst{
$^1$ Theoretical Physics Lab., Faculty of Knowledge Engineering,
Musashi Institute of Technology, Setagaya-ku, Tokyo 158-8557, Japan \\
$^2$ The Andrzej So{\l}tan Institute for Nuclear Studies, Ho\.{z}a 69, 00681, Warsaw, Poland}

\abst{ We argue that there is profound correspondence (the
nonextensive/dissipative correspondence - NexDC) between the
perfect nonextensive hydrodynamics and the usual dissipative
hydrodynamics which leads to simple expression for dissipative
entropy current.}

\begin{document}
\maketitle

We have recently proposed and discussed in detail
\cite{OsadaPRC77} a nonextensive hydrodynamical model, $q$-
hydrodynamics. It is based on the relativistic nonextensive
kinetic theory (as, for example, proposed in
works\cite{LavagnoPhysLettA301,LimaPhysRevLett86}), which
automatically accounts for all kinds of possible strong intrinsic
fluctuations and long-range correlations existing in systems of
quark and/or hadronic matter produced relativistic heavy-ion
collisions. It is expected to be in a kind of stationary
state\cite{KodamaJPhys31} rather than in exact thermal
equilibrium. It is described by nonextensive statistics
\cite{Tsallis1988} and characterized by the non-extensive
parameter $q$ . In this approach the {\it perfect
$q$-hydrodynamical equations} (for perfect, non-viscous in
$q$-language $q$-fluid) are given by
\begin{eqnarray}
{\cal T}_{q;\mu}^{\mu\nu} = \Big[ \varepsilon_q(T_q)
u_q^{\mu}u_q^{\nu} - P_q(T_q) \Delta_q^{\mu\nu} \Big]_{;\mu} =0.
\label{eq:q-equation_of_motion}
\end{eqnarray}
Here $\varepsilon_q(T_q)$, $P_q(T_q)$ and $u^{\mu}_q(x)$ are,
respectively, the nonextensive energy density and pressure (both
being functions of the none-extensive temperature $T_q$
\cite{OsadaPRC77}) and accompanying hydrodynamical flow four
vector, whereas $\Delta_q^{\mu\nu}\equiv g^{\mu\nu}-
u_q^{\mu}u_q^{\nu}$. Note that one can always decompose tensor
${\cal T}_q^{\mu\nu}$ by using another $4$-velocity field
$u^{\mu}(x)$ and obtain
\begin{eqnarray}
\left[\tilde{ \varepsilon} u^{\mu}u^{\nu}
-\tilde{P} \Delta^{\mu\nu} + 2W^{(\mu}u^{\nu)} +\pi^{\mu\nu} \right]_{;\mu} \!\!\!=0 ,
\label{eq:decomposition}
\end{eqnarray}
where (we denote $\delta u_q^{\mu} \equiv u_q^{\mu}-u^{\mu}$ and
$\Delta^{\mu\nu} \equiv g^{\mu\nu}- u^{\mu}u^{\nu}$)
\begin{subequations}
\label{eq:tilde_quantities}
\begin{eqnarray}
 &&\tilde{\varepsilon}=\varepsilon_q+3\Pi , \quad \tilde{P} = P_q +\Pi,  \label{eq:tilde_q12}\\
 &&{W}^{\mu}  
 = w_q[1+\gamma] ~\Delta^{\mu}_{\lambda} \delta u_q^{\lambda}, \label{eq:tilde_q3}  \\
&&
{\pi}^{\mu\nu} = \frac{W^{\mu} W^{\nu}}{ w_q [1+\gamma] ^2} +\Pi\Delta^{\mu\nu}
 =w_q~ \delta u_q^{<\mu}  \delta u_q^{\nu >}
 \label{eq:tilde_q4}
\end{eqnarray}
\end{subequations}
can be interpreted as, respectively, energy density
($\tilde{\varepsilon}$), pressure ($\tilde{P}$), energy or heat
flow vector ($W^{\mu}$) and shear pressure tensor ($\pi^{\mu\nu}$)
accompanying the field $u^{\mu}(x)$. Here $w_q \equiv
\varepsilon_q + P_q$, $\gamma \equiv u_{\mu}\delta u_q^{\mu} =
-\frac{1}{2}\delta u_{q\mu} \delta u_q^{\mu}$, $A^{( \mu}B^{\nu
)}\equiv \frac{1}{2}(A^{\mu}B^{\nu} +A^{\nu}B^{\mu})$,
$a^{<\mu}b^{ \nu >} \equiv [\frac{1}{2}(\Delta^{\mu}_{\lambda}
\Delta^{\nu}_{\sigma} + \Delta^{\mu}_{\sigma}
\Delta^{\nu}_{\lambda} ) -\frac{1}{3}\Delta^{\mu\nu}
\Delta_{\lambda\sigma} ] a^{\lambda}b^{\sigma}$ whereas
\begin{eqnarray}
\Pi  \equiv \frac{1}{3} w_q [\gamma^2+2\gamma] .\label{eq:bulk_pressure}
\end{eqnarray}
This last quantity can be regarded as a bulk pressure. The crucial
point of our work is assumption that there exists some temperature
$T$ and velocity field $\delta u_q^{\mu}$ satisfying the following
(which we call the {\it NexDC relations}):
\begin{eqnarray}
P(T)= P_q(T_q),\quad \varepsilon(T)= \varepsilon_q(T_q) +3\Pi.
\label{eq:Nex/diss}
\end{eqnarray}
Let $\varepsilon \equiv \varepsilon_{q=1}$ and $P \equiv P_{q=1}$
be the energy density and pressure (both functions of temperature
$T$) defined in the usual Boltzmann-Gibbs statistics (i.e., for
$q=1$). Using them one can transform equation
(\ref{eq:decomposition}) into following usual dissipative
hydrodynamical equation (or
$d$-hydrodynamics)\cite{Eckart1940,IsraelAnnPhys118,MurongaPRC69,
HeinzPRC73 }:
\begin{eqnarray}
\left[ \varepsilon(T) u^{\mu}u^{\nu}
\!-\!(P(T)+\Pi ) \Delta^{\mu\nu}
\!\!\!+\! 2 W^{(\mu} u^{\nu )}
\!+\!\pi^{\mu\nu} \right]_{;\mu} \!\!\!\!=0.
\label{eq:Nex/diss_equation}
\end{eqnarray}
This completes demonstration of our conjecture that perfect
$q$-hydrodynamics represented by Eq.
(\ref{eq:q-equation_of_motion}) is {\it equivalent} to
$d$-hydrodynamics represented by Eq. (\ref{eq:Nex/diss_equation}),
which is therefore its viscous counterpart. Notice that with bulk
pressure (\ref{eq:bulk_pressure}) and NexDC relations
(\ref{eq:Nex/diss}) one obtains the $q$-enthalpy,
\begin{eqnarray}
 \varepsilon_q(T_q)+P_q(T_q) =\frac{\varepsilon(T)+P(T)}{[1+\gamma]^2},
 \label{eq:enthalpy_correspondence}
\end{eqnarray}
which can be also used in definition of $\gamma$ because $w\equiv
Ts=\varepsilon+P$ and $1/(\gamma+1)= \sqrt{1-3\Pi/w}$ ($s$ is the
entropy density in the usual Boltzmann-Gibbs statistics). Notice
that in the NexDC one has the following relations:
\begin{eqnarray}
 W^{\mu} W_{\mu} \!= -3\Pi w,~
\pi^{\mu\nu}W_{\nu} \!= -2 \Pi W^{\mu},
~\pi_{\mu\nu}\pi^{\mu\nu} \!= 6\Pi^2.\quad
\label{eq:tensor_relations}
\end{eqnarray}

Let us consider now respective entropies. Dissipation is connected
with the production of entropy and in the usual approach
\cite{MurongaPRC69, IsraelAnnPhys118} the most general
off-equilibrium four-entropy current $\sigma^{\mu}$ is given by
\begin{eqnarray}
 \sigma^{\mu} = P(T)\beta^{\mu}
 + \beta_{\nu} ( T^{\mu\nu}_{\rm eq} +\delta T^{\mu\nu} ) +Q^{\mu},
  \label{eq:s_Israel}
\end{eqnarray}
where $\beta^{\mu}\equiv u^{\mu}/T$, $ {\cal T}^{\mu\nu}_{\rm eq}
\equiv \varepsilon(T)u^{\mu}u^{\nu} -P(T)\Delta^{\mu\nu}$, $\delta
{\cal T}^{\mu\nu}\equiv -\Pi\Delta^{\mu\nu} +W^{\mu}u^{\nu}
+W^{\nu}u^{\mu} +\pi^{\mu\nu}$ and where
$Q^{\mu}=Q^{\mu}\left(\delta \cal{T}^{\mu \nu}\right)$ is some
function which characterizes the off-equilibrium state. In the
case of the $q$-entropy current \cite{OsadaPRC77} the NexDC (i.e.,
Eqs. (\ref{eq:bulk_pressure}) and
(\ref{eq:enthalpy_correspondence})) leads to the following
off-equilibrium state:
\begin{eqnarray}
 Q^{\mu} \!\!&=&\!\! Q^{\mu}_{\chi}\equiv
 \chi \left[ su^{\mu} + \frac{W^{\mu}}{T}  \right],
 \label{eq:Q_expression}
\end{eqnarray}
(with $ \chi \equiv \frac{T}{T_q} \sqrt{1-\frac{3\Pi}{w}} - 1$)
which results in
\begin{eqnarray}
\sigma_{\chi}^{\mu} =
 su^{\mu}  + \frac{W^{\mu}}{T}
+\chi \left\{  su^{\mu}  + \frac{W^{\mu}}{T}  \right\} .
\end{eqnarray}
Notice that, because of the strict $q$-entropy conservation
assumed here, when using $Q^{\mu}=Q^{\mu}_{\chi}$  one always gets
$\sigma^{\mu}_{\chi;\mu}=0$. It means that, although there is no
production of $q$-entropy, there is some production of the usual
entropy, i.e., our $q$-system is really {\it dissipative} in the
usual meaning of this word.

Let us compare now the usual causal relativistic dissipative
theory as given by \cite{IsraelAnnPhys118,MurongaPRC69} with the
one emerging from our NexDC. The most general algebraic form of
$Q^{\mu}$, calculated up to the second order in the dissipative
flux, is given by~\cite{MurongaPRC69}
\begin{eqnarray}
  Q^{\mu}_{\rm 2nd} \!\!&=&\!\!
  \frac{\left[ -\beta_0 \Pi^2
                  +\beta_1 W_{\nu}W^{\nu}
                  -\beta_2 \pi_{\nu\lambda}\pi^{\nu\lambda} \right]}  {2T} u^{\mu}
    - \frac{\alpha_0\Pi W^{\mu}}{T} + \frac{\alpha_1\pi^{\mu\nu}W_{\nu}}{T},
\end{eqnarray}
where $\beta_{i=1,2,3}$ are the corresponding thermodynamic
coefficients for the, respectively, scalar, vector and tensor
dissipative contributions to the entropy current whereas
$\alpha_{i=0,1}$ are the corresponding viscous/heat coupling
coefficients. Correspondingly, in the NexDC one has
\begin{eqnarray}
  Q^{\mu}_{\rm 2nd}  \!\! &\to& \!\!
   \Gamma_{\rm 2nd}  ~su^{\mu}  +\Upsilon_{\rm 1st} \frac{W^{\mu}}{T},
\label{eqs:2nd_order_theory}
\end{eqnarray} where
 \begin{eqnarray}
 \Gamma_{\rm 2nd}
 \equiv -\frac{3\beta_1}{2}\Pi  -\frac{(\beta_0+6\beta_2)}{2w} \Pi^2 ,  
\quad
 \Upsilon_{\rm 1st}
 \equiv - (\alpha_0\ +2 \alpha_1) \Pi  .
 \end{eqnarray}
As one can see, in this case $Q^{\mu}$ can be expressed by
polynomials in the bulk pressure $\Pi$. Therefore, it is natural
to expect that the most general entropy current in the NexDC
approach has the following form:
\begin{eqnarray}
 Q^{\mu}_{\rm full} = \Gamma(\Pi) su^{\mu} +\Upsilon(\Pi)
 \frac{W^{\mu}}{T},
\end{eqnarray}
where $\Gamma ,\Upsilon$ are (in general infinite) series in
powers of the bulk pressure $\Pi$. In this sense the $Q^{\mu}_{\rm
full}$ can be regarded as being the {\it full order dissipative
current}.

In general one has entropy production/reduction, i.e.,
$\sigma^{\mu}_{;\mu} \ne 0$. However in the case when
$\Gamma(\Pi)= \Upsilon(\Pi)=\chi$ one has
$\sigma^{\mu}_{\chi;\mu}=0$ and therefore one can write the full
order dissipative entropy current as being equal to
\begin{eqnarray}
Q^{\mu}_{\rm full} =
( \chi + \xi )su^{\mu} +
( \chi - \xi )\frac{W^{\mu}}{T},
\end{eqnarray}
where $\Gamma$ and $\Upsilon$ are determined by $\chi\equiv
(\Gamma +\Upsilon)/2$ and $\xi\equiv (\Gamma-\Upsilon)/2$. Note
here that one always can express $\chi$ by $\kappa$ and $\gamma$:
\begin{eqnarray}
\chi= -\frac{\gamma}{(1+\gamma)(1+\kappa)} -\frac{\kappa}{1+\kappa} 
\label{eq:chi_expression}
\end{eqnarray}
where
$\kappa \equiv T_q/T-1$. The expression Eq. (\ref{eq:chi_expression})
suggests that one can get two possible solutions for $(\Gamma,\Upsilon)$
satisfying both $\chi\equiv(\Gamma +\Upsilon)/2$ and Eq. (\ref{eq:Q_expression}),
\begin{subequations}
\label{Q_general}
\begin{eqnarray}
 &&\frac{\Gamma}{2} \equiv \frac{T}{T_q}
\left(\sqrt{1-\frac{3\Pi}{w}}-1 \right),
    \quad \frac{\Upsilon}{2} \equiv \frac{T-T_q}{T_q}  \label{eq:Q_general1}\\
&&\hspace*{-12mm} \mbox{or} \nonumber \\
 && \frac{\Gamma}{2} \equiv  \frac{T-T_q}{T_q},
     \quad \frac{\Upsilon}{2} \equiv \frac{T}{T_q}  \left(\sqrt{1-\frac{3\Pi}{w}}-1
     \right).
      \label{eq:Q_general2}
\end{eqnarray}
\end{subequations}
Out of them only (\ref{eq:Q_general1}) is acceptable because only
for it $u_{\mu}Q^{\mu}_{\rm full} \le 0$ (i.e., entropy is maximal
in the equilibrium \cite{MurongaPRC69}, this is because
$(T-T_q)/T_q$ is always positive for $q\ge 1$ \cite{OsadaPRC77}).
In this way we finally arrive at the following possible expression
for the full order dissipative entropy current in the NexDC
approach:
\begin{eqnarray}
\sigma^{\mu}_{\rm full} 
\equiv
 su^{\mu} +\frac{W^{\mu}}{T}
-\frac{2T}{T_q}\left[~1- \sqrt{1-\frac{3\Pi}{w}} ~\right]su^{\mu} +
\frac{2(T-T_q)}{T_q}\frac{W^{\mu}}{T}.\quad
\label{eq:full_order_current}
\end{eqnarray}
Limiting ourselves to situations when $T/T_q \approx 1$ and
neglecting terms higher than ${\cal O}(3\Pi/w)^2$, one obtains
that
\begin{eqnarray}
Q^{\mu}_{\rm full}  \!\!&\approx&\!\!
\left[ -\left( \frac{3\Pi}{w} \right) -\frac{1}{4} \left( \frac{3\Pi}{w} \right)^2
\right] su^{\mu}. \label{eq:Q_chi_2nd}
\end{eqnarray}
Comparing Eqs.~(\ref{eqs:2nd_order_theory}) and
(\ref{eq:Q_chi_2nd}) one gets  $\beta_1 = \frac{2}{w},~\beta_0+
6\beta_2 = \frac{9}{2w},~\alpha_0+2\alpha_1=0$. Since in the
Israel-Stewart theory \cite{IsraelAnnPhys118} the relaxation time
$\tau$ is proportional to thermodynamical coefficients
$\beta_{0,1,2}$, it is naturally to assume that in our NexDC case
$\tau \propto 1/w$, i.e., it is proportional to the inverse of the
enthalpy.

To summarize, we have proposed to describe dissipative
hydrodynamics (at least partially) by using nonextensive
formulation of the usual perfect hydrodynamical model introducing
a nonextensive/dissipative correspondence (NexDC). As discussed in
more detail in \cite{OsadaPRC77} such model can be solved exactly
and when compared to the usual hydrodynamical approach it reveals
terms which can be interpreted as due to dissipative effects. They
can be therefore expressed by the single parameter of the theory
used, namely the nonextensivity parameter $q$. We have used this
finding to propose a possible full order expression for the
dissipative entropy current $\sigma^{\mu}_{{\rm full}}$.

\section*{Acknowledgements}
Partial support (GW) of the Ministry of Science and Higher
Education under contracts 1P03B02230 and  CERN/88/2006 is
acknowledged.

\end{document}